\documentclass[12pt]{article}
\pdfoutput=1

\usepackage{epsfig}
\usepackage{psfrag}
\usepackage{latexsym}
\usepackage{indentfirst}
\usepackage{fancyhdr}
\usepackage{dsfont}
\usepackage{amssymb}
\usepackage{amsmath}
\usepackage{amsfonts}
\usepackage{pifont}
\usepackage{cite}
\usepackage{bbold}
\usepackage{color}
\usepackage{colordvi}
\usepackage{fancybox}
\usepackage[footnotesize]{caption2}
\usepackage{graphicx}
\usepackage[center,footnotesize,hang]{subfigure}
\usepackage{bbm}
\usepackage{url}
\usepackage{multirow}
\usepackage{mathrsfs}
\usepackage{array}
\usepackage{arydshln}
\usepackage{ulem}
\usepackage{enumitem}
\newcommand{\PreserveBackslash}[1]{\let\temp=\\#1\let\\=\temp}
\newcolumntype{C}[1]{>{\PreserveBackslash\centering}p{#1}}
\newcolumntype{R}[1]{>{\PreserveBackslash\raggedleft}p{#1}}
\newcolumntype{L}[1]{>{\PreserveBackslash\raggedright}p{#1}}
\allowdisplaybreaks  

\newcommand{\bq}{\begin{eqnarray}}
\newcommand{\nq}{\end{eqnarray}}

\allowdisplaybreaks[1]

\def\simgt{\mathrel{\lower2.5pt\vbox{\lineskip=0pt\baselineskip=0pt
           \hbox{$>$}\hbox{$\sim$}}}}
\def\simlt{\mathrel{\lower2.5pt\vbox{\lineskip=0pt\baselineskip=0pt
           \hbox{$<$}\hbox{$\sim$}}}}

\newcommand{\be}{\begin{eqnarray}}
\newcommand{\ee}{\end{eqnarray}}

\def\bea{\begin{eqnarray}}
\def\eea{\end{eqnarray}}
\def\nn{\nonumber}

\begin{document}

\title{\hfill ~\\[0mm] \textbf{Neutrino Mixing from CP Symmetry}}

\date{}

\author{\\[1mm]Peng Chen\footnote{E-mail: {\tt pche@mail.ustc.edu.cn}}~,~~Chang-Yuan Yao\footnote{E-mail: {\tt phyman@mail.ustc.edu.cn}}~,~~Gui-Jun Ding\footnote{E-mail: {\tt dinggj@ustc.edu.cn}}\\ \\
\it{\small Department of Modern Physics, University of Science and
    Technology of China,}\\
  \it{\small Hefei, Anhui 230026, China}\\[4mm] }
\maketitle

\begin{abstract}
\noindent

The neutrino mass matrix has remnant CP symmetry expressed in terms of the lepton mixing matrix, and vice versa the remnant CP transformations allow us to reconstruct the mixing matrix. We study the scenario that all the four remnant CP transformations are preserved by the neutrino mass matrix. The most general parameterization of remnant CP transformations is presented. The lepton mixing matrix is completely fixed by the remnant CP, and its explicit form is derived. The necessary and sufficient condition for conserved Dirac CP violating phase is found. If the Klein four flavor symmetry generated by the postulated remnant CP transformations arises from a finite flavor symmetry group, the phenomenologically viable lepton flavor mixing would be the trimaximal pattern, both Dirac CP phase $\delta_{CP}$ and Majorana phase $\alpha_{31}$ are either $0$ or $\pi$ while another Majorana phase $\alpha_{21}$ is a rational multiple of $\pi$. These general results are confirmed to be true in the case that the  finite flavor symmetry group is $\Delta(6n^2)$.

\end{abstract}

\newpage

\section{\label{sec:introduction}Introduction}

The lepton flavor mixing is described by the Pontecorvo-Maki-Nakagawa-Sakata (PMNS) matrix which contains three mixing angles $\theta_{12}$, $\theta_{13}$ and $\theta_{23}$ and one Dirac CP phase $\delta_{CP}$~\cite{Agashe:2014kda}. There are two additional CP violating Majorana phases $\alpha_{21}$ and $\alpha_{31}$ which do not have any effect in neutrino oscillations, if neutrinos are Majorana particles. Many neutrino oscillation experiments have been performed all over the world and the values of mixing angles have been determined to a quite good precision~\cite{Gonzalez-Garcia:2014bfa},
\begin{equation}
0.270\leq\sin^2\theta_{12}\leq0.344,\quad 0.0188\leq\sin^2\theta_{13}\leq0.0251,\quad 0.385\leq\sin^2\theta_{23}\leq0.644\,
\end{equation}
at $3\sigma$ confidence level. Other global fits give rise to the similar results~\cite{Capozzi:2013csa,Forero:2014bxa}. On the other hand, the Dirac CP violating phase $\delta_{CP}$ is weakly constrained by the present neutrino oscillation data, and its $3\sigma$ interval is $[0, 2\pi]$~\cite{Gonzalez-Garcia:2014bfa,Capozzi:2013csa,Forero:2014bxa}. Therefore we don't know whether there is CP violation in the lepton sector at present. More recently, the T2K data show the hint that $\delta_{CP}$ is close to $-\pi/2$ (or $3\pi/2$)~\cite{Abe:2015awa}.

In the past years, much effort has been devoted to understanding the lepton mixing angles through the introduction of a finite discrete flavor symmetry $G_{f}$, which is subsequently broken down to a Klein subgroup $G_{\nu}=Z_2\times Z_2$ in the neutrino sector and an abelian group $G_{l}$ in the charged lepton sector~\cite{Altarelli:2010gt} under the assumption of Majorana neutrinos. In this approach, the lepton flavor mixing is completely determined by the embedding of the residual symmetries $G_{\nu}$ and $G_{l}$ into the original group $G_{f}$. A complete classification of the mixing patterns achievable in this way have been derived under the assumption that $G_{f}$ is finite~\cite{Fonseca:2014koa}. It turns out that the phenomenologically viable lepton mixing can only be of the trimaximal form and the Dirac phase $\delta_{CP}$ is trivial~\cite{Fonseca:2014koa}, and similar results are found even if neutrinos are Dirac particles~\cite{Yao:2015dwa}.

If the signal of $\delta_{CP}\simeq-\pi/2$ from T2K is further confirmed or $\delta_{CP}$ is measured to take a nontrivial value by future long baseline neutrino oscillation experiments~\cite{Adams:2013qkq,::2013kaa,Abe:2011ts}, the paradigm of discrete flavor symmetry would be disfavored, and we may have to resort to other new theoretical framework. Bottom-up analysis shows that the effective Majorana mass term of neutrino admits both Klein four flavor symmetry~\cite{Lam:2012ga} and four different CP symmetry transformations~\cite{Chen:2014wxa,Everett:2015oka}, and the flavor symmetry can be generated by performing two CP transformations in succession. Moreover, CP symmetry allows us to predict the values of Majorana CP phases which are not constrained by flavor symmetry at all. As a result, CP symmetry is a more general and more fundamental approach than flavor symmetry in explaining the lepton flavor mixing. Notice that only three of the four residual CP transformations of the neutrino mass term are independent. If two (or one) remnant CP transformations out of the original CP symmetry at high energy scale are preserved by the neutrino mass term in the charged lepton diagonal basis, the lepton mixing matrix would depend on one (or three) real parameters besides the parameters specifying the remnant CP transformations, and the explicit form of the mixing matrix has been presented in Ref.~\cite{Chen:2014wxa}. In this work, we shall consider the scenario that the maximal four CP transformations are preserved by the neutrino mass matrix. The most general parameterization of the CP transformations and the formula of constructing the PMNS matrix from the postulated CP transformations would be derived. Finally we would like to point that the phenomenological predictions and model building aspects of combining flavor symmetry with CP symmetry has been extensively discussed in the literatures~\cite{Mohapatra:2012tb, Feruglio:2012cw,King:2014rwa,Ding:2013bpa,Ding:2013hpa,Feruglio:2013hia,Luhn:2013lkn,Li:2013jya,Li:2014eia,Ding:2013nsa,Ding:2014ssa,Hagedorn:2014wha,Ding:2014ora,Li:2015jxa,DiIura:2015kfa,Ballett:2015wia}
in recent years. The interplay between flavor symmetry and CP symmetry have been comprehensively studied~\cite{Ecker:1981wv,Grimus:1995zi,Holthausen:2012dk,Chen:2014tpa}.

The remaining part of this paper is organized as follows. In section~\ref{sec:constructing_PMNS_matrix}, we present the most general parameterization of the full set of remnant CP transformations, and the explicit form of the PMNS matrix fixed by the remnant CP is derived. In section~\ref{sec:conditions_min_max_CP}, we investigate the conditions for minimal or maximal CP violating phases $\delta_{CP}$, $\alpha_{21}$ and $\alpha^{\prime}_{31}$. In section~\ref{sec:K4_finite}, we discuss the scenario that the Klein four flavor symmetry induced by the remnant CP transformations originates from a finite flavor symmetry group. Then the viable lepton mixing matrix can only be of the trimaximal form, both $\delta_{CP}$ and $\alpha_{31}$ are conserved, and $\alpha_{21}$ is a rational angle. As a concrete example, we apply our formalism to the case that the Klein four flavor symmetry generated by remnant CP is a subgroup of the $\Delta(6n^2)$ flavor symmetry in section~\ref{sec:Delta6n2}. Finally we present our conclusions in section~\ref{sec:conclusions}.

\section{\label{sec:constructing_PMNS_matrix}Reconstructing PMNS matrix from residual CP transformations}
We would assume in the present paper that neutrinos are Majorana particles. In the flavor basis, the Lagrangian describing the lepton masses reads
\begin{equation}
\label{eq:mass_Lagrange}\mathcal{L}_{mass}=-\overline{l}_{R}m_{l}l_{L}+\frac{1}{2}\nu^{T}_{L}C^{-1}m_{\nu}\nu_{L}+h.c.\,.
\end{equation}
Here $C$ is the charge-conjugation matrix, $l_{L}$ and $l_{R}$ denote the vectors of the three left and right-handed charged lepton fields, $\nu_{L}$ refers to the three left-handed neutrino fields, and $m_{l}\equiv\text{diag}\{m_{e}, m_{\mu}, m_{\tau}\}$. The neutrino mass matrix $m_{\nu}$ in the flavor basis, can be expressed via the mixing matrix $U_{PMNS}$ as
\begin{equation}
\label{eq:mnu_general}m_{\nu}=U^{\ast}_{PMNS}\text{diag}\left(m_1, m_2, m_3\right)U^{\dagger}_{PMNS}\,,
\end{equation}
where $m_1$, $m_2$ and $m_3$ are the light neutrino mass eigenvalues. The generic neutrino mass term in Eq.~\eqref{eq:mass_Lagrange} is invariant under both the CP transformations
\begin{equation}
\label{eq:res_CP}\nu_{L}(x)\stackrel{}{\longmapsto}iX_{j}\gamma^{0}C\bar{\nu}^{T}_{L}(x_P),\qquad X_{j}=U_{PMNS}\,d_j\,U^{T}_{PMNS},\quad j=1, 2, 3, 4
\end{equation}
and the flavor transformations
\begin{equation}
\label{eq:res_flavor}\hskip-0.4in \nu_{L}(x)\stackrel{}{\longmapsto} G_{i}\nu_{L}(x),\qquad G_{i}=U_{PMNS}\,d_i\,U^{\dagger}_{PMNS},\quad i=1, 2, 3, 4\,,
\end{equation}
where $x_{P}=(t,-\vec{x})$ and
\begin{eqnarray}
\nonumber&& d_1=\text{diag}\left(1,-1,-1\right),\qquad d_2=\text{diag}\left(-1,1,-1\right),\\
&&d_3=\text{diag}\left(-1,-1,1\right),\qquad d_4=\text{diag}\left(1,1,1\right)\,.
\end{eqnarray}
As has been shown in our previous work~\cite{Chen:2014wxa}, only three of the four remnant CP transformations are independent. The reason is that any one of the residual CP transformations can be generated by the remaining three via
\begin{equation}
\label{eq:CP_independent}X_{i}=X_{j}X^{\ast}_{m}X_{n},~~\quad i\neq j\neq m\neq n\,.
\end{equation}
The remnant flavor symmetry and remnant CP transformations are closely related with each other. The remnant flavor symmetry can be generated by performing two CP transformations as follows,
\begin{eqnarray}
\nonumber&&G_1=X_2X^{\ast}_3=X_3X^{\ast}_2=X_4X^{\ast}_1=X_1X^{\ast}_4,\\
\nonumber&&G_2=X_1X^{\ast}_3=X_3X^{\ast}_1=X_4X^{\ast}_2=X_2X^{\ast}_4,\\
\nonumber&&G_3=X_1X^{\ast}_2=X_2X^{\ast}_1=X_4X^{\ast}_3=X_3X^{\ast}_4,\\
\label{eq:CP_flavor}&&G_4=X_1X^{\ast}_1=X_2X^{\ast}_2=X_3X^{\ast}_3=X_4X^{\ast}_4=1\,.
\end{eqnarray}
Furthermore, we can straightforwardly see that $X_i$ and $G_j$ fulfill the following relation
\begin{eqnarray}
X_i G_j^{\ast} X_i^{\ast}=U_{PMNS}\,d_j\,U^{\dagger}_{PMNS}=G_j~~\mathrm{for}~~ i,j=1,2,3,4\,.
\label{eq:XGcond}
\end{eqnarray}
This means that the residual flavor symmetry and residual CP symmetry should generally commute with each other in the neutrino sector. Given the experimentally measured mixing matrix $U_{PMNS}$, the CP transformation matrix $X_{i}$ can be easily fixed by Eq.~\eqref{eq:res_CP}. Inversely, we shall show that $U_{PMNS}$ can be deduced from any well-defined four CP transformations.

The scenario of one or two remnant CP transformations preserved by the neutrino mass matrix have been investigated in our previous work~\cite{Chen:2014wxa}. In the following, we shall consider the scenario that an original CP symmetry at high energy scale is broken down to four remnant CP transformations $X_{i}~(i=1,2,3,4)$ in the neutrino sector by some scalar fields. Notice that if three CP transformations are preserved by the neutrino mass matrix, there are still four residual CP transformations since the fourth one can be generated via Eq.~\eqref{eq:CP_independent}. Here we shall not consider how the remnant CP transformations $X_{i}~(i=1,2,3,4)$ are dynamically achieved, since the lepton flavor mixing is completely fixed by the remnant CP and it is independent of the specific breaking mechanisms. $X_{i}~(i=1,2,3,4)$ can be treated as remnant CP transformations only if the following consistency conditions are fulfilled,
\begin{eqnarray}
\nonumber&&X_{i}=X^{T}_{i},\quad  X_{i}X^{\ast}_{j}=X_{j}X^{\ast}_{i}=X_{m}X^{\ast}_{n}=X_{n}X^{\ast}_{m},\\
\label{eq:consistency_cond}&&\left( X_{i}X^{\ast}_{j}\right)^2=1,\quad X_{i}X^{\ast}_{j}\ne\pm1,~~\mathrm{for}~~ i\neq j\neq m\neq n\,.
\end{eqnarray}
In this case, a full Klein four flavor symmetry is generated with element of the form $X_{i}X^{\ast}_{j}~(i\neq j)$. First of all, we present the parameterization of the whole set of CP transformations in the following. Using the freedom in redefining the phases of the charged lepton fields, one column of the PMNS matrix can always be set to be real and written as
\begin{equation}
\label{eq:v1}v_1=\left(
\begin{array}{c}
\cos\varphi            \\
\sin\varphi\cos\phi    \\
\sin\varphi\sin\phi    \\
\end{array}
\right)\,,
\end{equation}
where both $\varphi$ and $\phi$ are real parameters in the interval of $[0, 2\pi)$. As a result, one of the remnant flavor symmetry is given by
\begin{equation}
G_1=2v_1v^{\dagger}_1-1=2v_1v_1^T-1\,.
\end{equation}
Obviously $G_{1}$ is a real matrix in our working basis, and then Eq.~\eqref{eq:XGcond} implies that
\begin{equation}
\left[X_{i},G_{1}\right]=0,\quad \mathrm{for}\quad i=1,2,3,4,
\end{equation}
Therefore $v_1$ should be a eigenvector of $X_{i}$ as well. As the remnant CP transformations $X_{i}$ are symmetric and unitary matrices, the eigenvalues of $X_{i}$ must be of unit modulus and the eigenvectors of $X_{i}$ can be chosen to be real. Following Ref.~\cite{Chen:2014wxa}, we can conveniently parameterize the remnant CP transformation $X_{i}$ in terms of its eigenvalues and eigenvectors. For instance, we can write $X_{1}$ as
\begin{equation}
X_{1}= e^{i\kappa_1}v_1v^T_1+e^{i\kappa_2}v_2v^T_2+e^{i\kappa_3}v_3v^T_3\,,
\label{eq:X1}
\end{equation}
where $\kappa_1$, $\kappa_2$ and $\kappa_3$ are real, $v_1$, $v_2$ and $v_3$ are a set of most general real orthonormal vectors, and $v_2$ and $v_3$ are given by
\begin{eqnarray}
v_2=\left(
\begin{array}{c}
\sin\varphi\cos\rho   \\
-\sin\phi\sin\rho-\cos\varphi\cos\phi\cos\rho   \\
\cos\phi\sin\rho-\cos\varphi\sin\phi\cos\rho
\end{array}\right)\,, \quad
v_3=\left(
\begin{array}{c}
\sin\varphi\sin\rho     \\
\sin\phi\cos\rho-\cos\varphi\cos\phi\sin\rho    \\
-\cos\phi\cos\rho-\cos\varphi\sin\phi\sin\rho
\end{array}\right)\,.
\end{eqnarray}
From the expression of $X_{1}$ in Eq.~\eqref{eq:X1} and the relation of Eq.~\eqref{eq:CP_flavor}, $X_{4}$ can be constructed directly
\begin{eqnarray}
X_{4}=G_1X_1=e^{i\kappa_1}v_1v^T_1-e^{i\kappa_2}v_2v^T_2-e^{i\kappa_3}v_3v^T_3\,.
\label{eq:X4}
\end{eqnarray}
In a similar way, we can parameterize $X_{2}$ and $X_{3}$ as follows
\begin{eqnarray}
X_{2}&=&e^{i\lambda_1}v_1v^T_1+e^{i\lambda_2}w_2w^T_2+e^{i\lambda_3}w_3w^T_3\,,\nn\\
X_{3}&=&e^{i\lambda_1}v_1v^T_1-e^{i\lambda_2}w_2w^T_2-e^{i\lambda_3}w_3w^T_3\,,\label{eq:X23}
\end{eqnarray}
where $v_1$, $w_2$ and $w_3$ also form a set of real orthonormal vectors, and $w_2$ and $w_3$ take the form
\begin{equation}
w_2=\cos\xi v_2-\sin\xi v_3,\qquad w_3=\sin\xi v_2+\cos\xi v_3\,,
\end{equation}
The residual CP transformations $X_1$, $X_2$, $X_3$ and $X_4$ given by Eqs.~(\ref{eq:X1}, \ref{eq:X4}, \ref{eq:X23}) have to satisfy the consistency conditions in Eq.~\eqref{eq:consistency_cond}. We can straightforwardly obtain the expression for the product $X_iX^{\ast}_j$ as follows
\begin{eqnarray}
X_{1}X_{1}^{\ast}&=&v_1v^{T}_1+v_2v^{T}_2+v_2v^{T}_3=1\,,\nn\\
X_{1}X_{2}^{\ast}&=&e^{i(\kappa_1-\lambda_1)}v_1v^T_1+c_{22}v_2v_2^T+c_{33}v_3v_3^T+c_{23}v_2v_3^T+c_{32}v_3v_2^T\,,\nn\\
X_{1}X_{3}^{\ast}&=&e^{i(\kappa_1-\lambda_1)}v_1v^T_1-c_{22}v_2v_2^T-c_{33}v_3v_3^T-c_{23}v_2v_3^T-c_{32}v_3v_2^T\,,\nn\\
X_{1}X_{4}^{\ast}&=&v_1v_1^T-v_2v_2^T-v_3 v_3^T\,,\nn\\
X_{2}X_{1}^{\ast}&=&e^{i(\lambda_1-\kappa_1)}v_1v^T_1+c_{22}^*v_2v_2^T+c_{33}^*v_3v_3^T+c_{32}^*v_2v_3^T+c_{23}^*v_3v_2^T\,,\nn\\
X_{2}X_{2}^{\ast}&=&v_1v^{T}_1+v_2v^{T}_2+v_3v^{T}_3=1\,,\nn\\
X_{2}X_{3}^{\ast}&=&v_1v_1^T-v_2v_2^T-v_3v_3^T\,,\nn\\
X_{2}X_{4}^{\ast}&=&e^{i(\lambda_1-\kappa_1)}v_1v^T_1-c_{22}^*v_2v_2^T-c_{33}^*v_3v_3^T-c_{32}^*v_2v_3^T-c_{23}^*v_3v_2^T\,,\nn\\
X_{3}X_{1}^{\ast}&=&e^{i(\lambda_1-\kappa_1)}v_1v^T_1-c_{22}^*v_2v_2^T-c_{33}^*v_3v_3^T-c_{32}^*v_2v_3^T-c_{23}^*v_3v_2^T\,,\nn\\
X_{3}X_{2}^{\ast}&=&v_1v_1^T-v_2v_2^T-v_3v_3^T\,,\nn\\
X_{3}X_{3}^{\ast}&=&v_1v^{T}_1+v_2v^{T}_2+v_3v^{T}_3=1\,,\nn\\
X_{3}X_{4}^{\ast}&=&e^{i(\lambda_1-\kappa_1)}v_1v^T_1+c_{22}^*v_2v_2^T+c_{33}^*v_3v_3^T+c_{32}^*v_2v_3^T+c_{23}^*v_3v_2^T\,,\nn\\
X_{4}X_{1}^{\ast}&=&v_1v_1^T-v_2v_2^T-v_3v_3^T\,,\nn\\
X_{4}X_{2}^{\ast}&=&e^{i(\kappa_1-\lambda_1)}v_1v^T_1-c_{22}v_2v_2^T-c_{33}v_3v_3^T-c_{23}v_2v_3^T-c_{32}v_3v_2^T\,,\nn\\
X_{4}X_{3}^{\ast}&=&e^{i(\kappa_1-\lambda_1)}v_1v^T_1+c_{22}v_2v_2^T+c_{33}v_3v_3^T+c_{23}v_2v_3^T+c_{32}v_3v_2^T\,,\nn\\
X_{4}X_{4}^{\ast}&=&v_1v_1^T+v_2v_2^T+v_3v_3^T=1\,,
\end{eqnarray}
where we have defined
\begin{eqnarray}
c_{22}&\equiv&\left(e^{-i\lambda_2}\cos^2\xi+e^{-i\lambda_3}\sin^2\xi\right)e^{i\kappa_2}\,,\nn\\
c_{33}&\equiv&\left(e^{-i\lambda_2}\sin^2\xi+e^{-i\lambda_3}\cos^2\xi\right)e^{i\kappa_3}\,,\nn\\
c_{23}&\equiv&\left(e^{-i\lambda_3}-e^{-i\lambda_2}\right)e^{i\kappa_2}\cos\xi\sin\xi\,,\nn\\
c_{32}&\equiv&\left(e^{-i\lambda_3}-e^{-i\lambda_2}\right)e^{i\kappa_3}\cos\xi\sin\xi\,.
\end{eqnarray}
The equations in the first line of Eq.~\eqref{eq:consistency_cond} give rise to the following constraints
\begin{equation}
e^{i(\kappa_1-\lambda_1)}=e^{i(\lambda_1-\kappa_1)},\quad c_{22}^*=c_{22}, \quad c_{33}^*=c_{33}, \quad c_{23}^*=c_{32}\,.\label{eq:constraint1}
\end{equation}
Furthermore, the requirement of $(X_iX^{\ast}_j)^2=1$ in the second line of Eq.~\eqref{eq:consistency_cond} leads to
\begin{eqnarray}
&& e^{2 i(\kappa_1-\lambda_1)}=1,\quad c_{22}^2+c_{23}c_{32}=1,\quad c_{33}^2+c_{23}c_{32}=1, \nn\\
&&\qquad \left(c_{22}+c_{33}\right)c_{23}=0,\qquad
\left(c_{22}+c_{33}\right)c_{32}=0\,.
\label{eq:constraint2}
\end{eqnarray}
The solutions to the above set of equations in Eqs.~(\ref{eq:constraint1},\ref{eq:constraint2}) are given by

\begin{equation}
e^{i\lambda_1}=\mp e^{i\kappa_1},\quad e^{i\lambda_2}=\mp\frac{e^{i\kappa_2}\cos^2\xi+e^{i\kappa_3}\sin^2\xi}{\left|e^{i\kappa_2}\cos^2\xi+ e^{i\kappa_3}\sin^2\xi\right|},\quad e^{i\lambda_3}=\pm\frac{e^{i\kappa_2}\sin^2\xi+e^{i\kappa_3}\cos^2\xi}{\left|e^{i\kappa_2}\sin^2\xi+e^{i\kappa_3}\cos^2\xi\right|}\,,
\end{equation}
where the ``$\mp$'' signs in $e^{i\lambda_1}$ can be chosen independently. As a consequence, the parameters $c_{22}$, $c_{23}$, $c_{32}$ and $c_{33}$ are simplified into
\begin{eqnarray}
c_{22}&=&\mp\frac{\cos2\xi}{\left|e^{i\kappa_2}\cos^2\xi+e^{i\kappa_3}\sin^2\xi\right| },\\
c_{23}&=&\pm\frac{\cos\left(\frac{\kappa_2-\kappa_3}{2}\right)e^{i\frac{\kappa_2-\kappa_3}{2}} \sin2\xi }{\left|e^{i\kappa_2}\cos^2\xi + e^{i\kappa_3}\sin^2\xi \right|},\\
c_{32}&=&\pm\frac{\cos\left(\frac{\kappa_2-\kappa_3}{2}\right)e^{-i\frac{\kappa_2-\kappa_3}{2}} \sin2\xi}{\left|e^{i\kappa_2}\cos^2\xi+e^{i\kappa_3}\sin^2\xi\right|},\\
c_{33}&=&\pm\frac{\cos2\xi}{\left|e^{i\kappa_2}\cos^2\xi+e^{i\kappa_3}\sin^2\xi\right|}\,.
\end{eqnarray}
Since the above solutions give rise to the same set of residual CP transformations $X_{1,2,3,4}$ up to an irrelevant overall ``$-1$'' factor, without loss of generality, we could choose
\begin{equation}
e^{i\lambda_1}=-e^{i\kappa_1},\quad e^{i\lambda_2}=-\frac{e^{i\kappa_2}\cos^2\xi+e^{i\kappa_3}\sin^2\xi}{\left|e^{i\kappa_2}\cos^2\xi+ e^{i\kappa_3}\sin^2\xi\right|},\quad e^{i\lambda_3}=\frac{e^{i\kappa_2}\sin^2\xi+e^{i\kappa_3}\cos^2\xi}{\left|e^{i\kappa_2}\sin^2\xi+e^{i\kappa_3}\cos^2\xi\right|}\,.
\end{equation}
A remnant Klein four flavor symmetry $K_4\equiv\left\{1, G_1, G_2, G_3\right\}$ can be generated by performing two CP transformations, and the three nontrivial residual flavor symmetry transformations $G_i$ for $i=1, 2, 3$ can be written into
\begin{eqnarray}
G_{1} &=& X_{1}X_{4}^{\ast} =
\left(v_1, v_2, v_3\right)
\left(\begin{array}{ccc}
1   ~&~   0   ~&~    0     \\
0   ~&~  -1   ~&~    0     \\
0   ~&~   0   ~&~    -1
\end{array}
\right)
\left(\begin{array}{c}
v_1^T        \\
v_2^T        \\
v_3^T
\end{array}
\right)\,,
\nonumber\\
G_{2} &=& X_{1}X_{3}^{\ast} =
\left(v_1, v_2, v_3\right)
\left(\begin{array}{ccc}
-1  &   0       &       0     \\
0    &  -c_{22}    &  -c_{23}     \\
0    &  -c_{32}    &  -c_{33}
\end{array}
\right)
\left(
\begin{array}{c}
v_1^T        \\
v_2^T        \\
v_3^T
\end{array}
\right)\,,\nonumber\\
G_{3}&=&X_{1}X_{2}^{\ast} =
\left(v_1, v_2, v_3\right)
\left(\begin{array}{ccc}
-1  ~&~   0       ~&~       0     \\
0   ~&~  c_{22}   ~&~  c_{23}     \\
0   ~&~  c_{32}   ~&~  c_{33}
\end{array}
\right)
\left(\begin{array}{c}
v_1^T         \\
v_2^T         \\
v_3^T
\end{array}
\right)\,.
\label{eq:G123}
\end{eqnarray}
We notice that the matrix
\begin{equation}
\left(\begin{array}{ccc}
-1   &   0    &    0     \\
0    &  c_{22}   &  c_{23}   \\
0    &  c_{32}   &  c_{33}
\end{array}
\right)
\end{equation}
can be diagonalized by a $3\times3$ matrix $\Theta_{3\times3}$ with
\begin{equation}
\Theta_{3\times3}=\left(
\begin{array}{ccc}
1   ~&~  0   &  0     \\
0   ~&~ \cos\theta   &  \sin\theta e^{i\frac{\kappa_2-\kappa_3}{2}}  \\
0   ~&~ -\sin\theta e^{-i\frac{\kappa_2-\kappa_3}{2}}  &  \cos\theta
\end{array}
\right)\,,
\end{equation}
where the rotation angle $\theta$ fulfills
\begin{eqnarray}
\tan2\theta&=&\cos\left(\frac{\kappa_2-\kappa_3}{2}\right)\tan2\xi\,,\nn\\
\cos2\theta&=&\frac{\cos2\xi}{\sqrt{\cos^22\xi+\cos^2\left(\frac{\kappa_2-\kappa_3}{2} \right)\sin^22\xi}}\,.
\label{eq:theta_value}
\end{eqnarray}
In other words, we have
\begin{equation}
\Theta_{3\times3}^{\dagger}\left(
\begin{array}{ccc}
-1   &  0    &    0     \\
0    &  c_{22}   &  c_{23}     \\
0    &  c_{32}   &  c_{33}
\end{array}\right)\Theta_{3\times3}=\mathrm{diag}(-1, -1, 1)\,. \label{eq:Cmatrix_Dia}
\end{equation}
We can introduce three column vectors $\mathscr{V}_1$, $\mathscr{V}_2$ and $\mathscr{V}_3$ with
\begin{equation}
\left(\mathscr{V}_1, \mathscr{V}_2, \mathscr{V}_3\right)=\left(v_1, v_2, v_3\right)\Theta_{3\times3}\,.
\end{equation}
Then the residual flavor symmetries $G_1$, $G_2$ and $G_3$ can be written as
\begin{eqnarray}
G_{1}=2\mathscr{V}_1\mathscr{V}_1^\dagger-1\,,\qquad G_{2}=2\mathscr{V}_2\mathscr{V}_2^\dagger-1\,, \qquad G_{3}=2\mathscr{V}_3\mathscr{V}_3^\dagger-1\,.\label{eq:Gi}
\end{eqnarray}
The three vectors $\mathscr{V}_1$, $\mathscr{V}_2$ and $\mathscr{V}_3$ are the unique eigenvectors of $G_1$, $G_2$ and $G_3$ respectively with eigenvalue $+1$, and they provide the three columns of the PMNS matrix up to permutations and phases. Therefore the PMNS matrix is determined to be
\begin{equation}
U_{PMNS}=\left(\mathscr{V}_1, \mathscr{V}_2, \mathscr{V}_3\right)\text{diag}(e^{i\gamma_1}, e^{i\gamma_2}, e^{i\gamma_3})P=\left(v_1, v_2, v_3\right)\Theta_{3\times3}\,\text{diag}(e^{i\gamma_1}, e^{i\gamma_2}, e^{i\gamma_3})P\,,
\end{equation}
where $P$ is an arbitrary permutation matrix since the light neutrino masses are unconstrained in the present framework. The phases $\gamma_1$, $\gamma_2$ and $\gamma_3$ are further subject to the constraints from  the postulated residual CP transformations as $U^{\dagger}_{PMNS}X_iU^{\ast}_{PMNS}=\text{diag}(\pm1, \pm1, \pm1)$ in Eq.~\eqref{eq:res_CP}, and they are fixed to be
\begin{equation}
\label{cond1}\text{diag}\left(e^{i\gamma_1}, e^{i\gamma_2}, e^{i\gamma_3}\right)=\text{diag}\left(e^{i\kappa_1/2}, e^{i\kappa_2/2}, e^{i\kappa_3/2}\right)PKP^{T}\,,
\end{equation}
where $K$ is a diagonal phase matrix with entries $\pm1$ or $\pm i$, it render the light neutrino to be non-negative. As a result, the PMNS matrix is completely determined by the residual CP transformations to be
\begin{eqnarray}
U_{PMNS}&=&\left(v_1, v_2, v_3\right)\Theta_{3\times3}\,\text{diag}(e^{i\kappa_1/2}, e^{i\kappa_2/2}, e^{i\kappa_3/2})PK\nn\\
&=&\left(\begin{array}{ccc}
\cos\varphi    ~&~ \sin\varphi  ~&~   0    \\
\sin\varphi\cos\phi   ~&~  -\cos\varphi\cos\phi    ~&~  \sin\phi  \\
\sin\varphi\sin\phi   ~&~  -\cos\varphi\sin\phi    ~&~  -\cos\phi  \end{array}\right)
\left(\begin{array}{ccc}
1   ~&~   0   ~&~    0           \\
0   ~&~  \cos\rho  ~&~  \sin\rho     \\
0   ~&~  -\sin\rho ~&~  \cos\rho
\end{array}\right)
\left(\begin{array}{ccc}
e^{i\kappa_1/2}   ~&~         0          ~&~       0        \\
0                 ~&~   e^{i\kappa_2/2}  ~&~       0        \\
0                 ~&~         0          ~&~  e^{i\kappa_3/2}
\end{array}\right)\nonumber\\
&&\times\left(\begin{array}{ccc}
1   ~&~  0   ~&~  0     \\
0   ~&~ \cos\theta   ~&~  \sin\theta  \\
0   ~&~ -\sin\theta  ~&~  \cos\theta
\end{array}\right)PK\,,
\label{eq:UPMNS_final}
\end{eqnarray}
where the contribution from the matrix $K$ can only possibly shift the Majorana phases by $\pi$. The different arrangements of rows and columns can be related by parameter redefinition in the present framework. Without loss of generality, we choose the column permutation $P$ in Eq.~\eqref{eq:UPMNS_final} to be
\begin{equation}
\label{eq:P_example}P=\left(\begin{array}{ccc}
0   &  0  &  1  \\
1   &  0  &  0  \\
0   &  1  &  0
\end{array}\right)
\end{equation}
for illustration. The predictions for the lepton mixing parameters can be straightforwardly read out as follows:
\begin{eqnarray}
\sin^2\theta_{13}&=&\cos^2\varphi, \quad \sin^2\theta_{12}=\frac{1}{2}\left(1-\cos2\rho\cos2\theta+\cos\frac{\kappa^\prime_2-\kappa^\prime_3}{2}\sin2\rho\sin2\theta\right),\nn\\
\sin^2\theta_{23}&=&\cos^2\phi,\quad \tan\delta_{CP}=\frac{\sin(\frac{\kappa^\prime_2-\kappa^\prime_3}{2})}{\sin2\rho\cot2\theta+\cos(\frac{\kappa^\prime_2-\kappa^\prime_3}{2})\cos2\rho},\nn\\
\tan\alpha_{21}&=&-\frac{2\sin(\kappa^\prime_2-\kappa^\prime_3)\sin^22\rho\cos2\theta+2\sin(\frac{\kappa^\prime_2-\kappa^\prime_3}{2})\sin4\rho\sin2\theta}
{(3\cos^22\rho-1)\sin^22\theta+\cos(\kappa^\prime_2-\kappa^\prime_3)\sin^22\rho(\cos^22\theta+1)+\cos(\frac{\kappa^\prime_2-\kappa^\prime_3}{2})\sin4\rho\sin4\theta},\nn\\
\tan\alpha^{\prime}_{31}&=&-\frac{2\cos^2\rho\cos^2\theta\sin\kappa^\prime_2+2\sin^2\rho\sin^2\theta\sin\kappa^\prime_3-\sin2\rho\sin2\theta\sin(\frac{\kappa^\prime_2+\kappa^\prime_3}{2})}
{2\cos^2\rho\cos^2\theta\cos\kappa^\prime_2+2\sin^2\rho\sin^2\theta\cos\kappa^\prime_3-\sin2\rho\sin2\theta\cos(\frac{\kappa^\prime_2+\kappa^\prime_3}{2})}\,,\label{eq:mixing_para}
\end{eqnarray}
where $\alpha^{\prime}_{31}=\alpha_{31}-2\delta_{CP}$, $\delta_{CP}$ is the Dirac CP violating phase in standard parameterization, $\alpha_{21}$ and $\alpha_{31}$ are the Majorana CP phases~\cite{Agashe:2014kda}. Since $e^{i\kappa_1}$ can be factorized out as an overall phase of the PMNS matrix in Eq.~\eqref{eq:UPMNS_final}, all the mixing parameters only depend on the phase differences $\kappa_2^\prime\equiv\kappa_2-\kappa_1$ and $\kappa_3^\prime\equiv\kappa_3-\kappa_1$. Notice that $\theta_{12}$ and $\theta_{13}$ only relate to $\phi$ and $\varphi$ respectively. Moreover, the well-known Jarlskog invariant~\cite{Jarlskog:1985ht} is
\begin{equation}
\label{eq:Jarlskog}J_{CP}=\frac{1}{4}\sin(\frac{\kappa^\prime_2-\kappa^\prime_3}{2})\cos\varphi\sin^2\varphi\sin2\phi\sin2\theta\,.
\end{equation}
Notice that only the overall sign of $J_{CP}$ could possibly change for other permutation $P$ distinct from that in Eq.~\eqref{eq:P_example}.

\section{\label{sec:conditions_min_max_CP}Conditions for vanishing and maximal CP violation}

\begin{table}[t!]
\renewcommand{\arraystretch}{1.2}
\begin{center}
\footnotesize
\begin{tabular}{|c|c|c|c|}\hline\hline
\multicolumn{4}{|c|}{\texttt{Minimal $\delta_{CP}$ with $\sin\delta_{CP}=0$}}\\  \hline

\texttt{Conditions}  & $\sin^2\theta_{12}$  &  $\tan\alpha_{21}$  &   $\tan\alpha_{31}$  \\\hline

 & & &   \\[-0.14in]

$\xi=n\pi/2$  &   $\sin^2\left(\rho+\xi\right)$   &   $(-1)^{n-1}\tan(\kappa^{\prime}_2-\kappa^{\prime}_3)$  &   $\left\{
\begin{array}{cr}
-\tan\kappa^{\prime}_3,&~  \text{n is odd}\\
-\tan\kappa^{\prime}_2,&~  \text{n is even}
\end{array}
\right.$  \\
 & & &   \\[-0.14in] \hline

 & & &   \\[-0.14in]

$\kappa_3=\kappa_2+n\pi$  &   $\left\{
\begin{array}{cc}
\frac{1}{2}\left[1-\text{sign}(\cos2\xi)\cos2\rho\right],&~ \text{n is odd} \\
\sin^2(\rho+\xi),&~ \text{n is even}
\end{array}
\right.$   &   0  &  $-\tan\kappa^{\prime}_2$  \\
 & & &   \\[-0.14in]\hline\hline

\multicolumn{4}{|c|}{\texttt{Minimal $\alpha_{21}$ with $\sin\alpha_{21}=0$}}\\  \hline

\texttt{Conditions}  & $\sin^2\theta_{12}$  &  $\tan\delta_{CP}$  &   $\tan\alpha_{31}$  \\\hline

 & & &   \\[-0.14in]

$\kappa_3=\kappa_2+n\pi$  &   $\left\{\begin{array}{cc}
\frac{1}{2}\left[1-\text{sign}(\cos2\xi)\cos2\rho\right],&~ \text{n is odd} \\
\sin^2(\rho+\xi), & ~ \text{n is even}\end{array}\right.$  &   0  & $-\tan\kappa^{\prime}_2$  \\
 & & &   \\[-0.14in] \hline

 & & &   \\[-0.14in]

$\rho=n\pi/2$  &   $\frac{1}{2}\Big[1-\frac{(-1)^n\cos2\xi}
{\sqrt{\cos^22\xi+\cos^2\left(\frac{\kappa^\prime_2-\kappa^\prime_3}{2}\right)\sin^22\xi}}\Big]$ &   $(-1)^n\tan\frac{\kappa_2^\prime-\kappa_3^\prime}{2}$   &   $\left\{\begin{array}{cc}
-\tan\kappa^{\prime}_3,  &~    \text{n is odd} \\
-\tan\kappa^{\prime}_2,  &~    \text{n is even}
\end{array}\right.$  \\
 & & &   \\[-0.14in]\hline

 & & &   \\[-0.14in]

$\rho=-\xi+n\pi$   &  $\frac{1}{2}\left(1-\sqrt{\cos^22\xi+\cos^2\frac{\kappa_2^\prime-\kappa_3^\prime}{2}\sin^22\xi}\right)$ &  $-\cot\frac{\kappa_2^\prime-\kappa_3^\prime}{2}\sec2\xi$   &  $-\frac{\cos^2\xi\sin\kappa_2^\prime+\sin^2\xi\sin\kappa_3^\prime}{\cos^2\xi\cos\kappa_2^\prime+\sin^2\xi\cos\kappa_3^\prime}$  \\
 & & &   \\[-0.14in]\hline\hline

\end{tabular}
\end{center}
\renewcommand{\arraystretch}{1.0}
\caption{\label{tab:minimal_deltaCP_alpha_21}
The conditions for trivial Dirac phase $\delta_{CP}$ and trivial Majorana phase $\alpha_{21}$ with $\sin\delta_{CP}=0$ and $\sin\alpha_{21}=0$ respectively, where $n$ is an integer. The predictions for the mixing parameters are listed here. Note that we have $\sin^2\theta_{13}=\cos^2\varphi$ and $\sin^2\theta_{23}=\cos^2\phi$ for the PMNS matrix in Eq.~\eqref{eq:UPMNS_final} with the permutation matrix $P$ given by Eq.~\eqref{eq:P_example}.}
\end{table}

From the analytical expressions of the mixing parameters shown in  Eq.~\eqref{eq:mixing_para}, we can straightforwardly derive the conditions for minimal or maximal CP phases. Firstly, we find that the Jarlskog invariant $J_{CP}$ in Eq.~\eqref{eq:Jarlskog} is vanishing such that the Dirac CP violating phase $\delta_{CP}$ would be conserved if and only if
\begin{equation}
\label{eq:minimal_deltaCP}\xi=0,~\frac{1}{2}\pi,~\pi,~\frac{3}{2}\pi,~~\mathrm{or}~~ \kappa_3=\kappa_2,~~\mathrm{or}~~ \kappa_3=\kappa_2\pm\pi\,.
\end{equation}
The resulting predictions for the remaining mixing parameters are collected in Table~\ref{tab:minimal_deltaCP_alpha_21} for the column permutation $P$ of Eq.~\eqref{eq:P_example}. Note that one element of the PMNS matrix would be vanishing such that the value of $\delta_{CP}$ can not be fixed uniquely in case of $\phi=n\pi/2$ or $\varphi=n\pi/2$ where $n=0, 1, 2, 3$. It is remarkable that $\delta_{CP}$ is always trivial with $\sin\delta_{CP}=0$ for any value of the permutation $P$ once the conditions in Eq.~\eqref{eq:minimal_deltaCP} are fulfilled. In other words, the Dirac CP would be violated if the parameters $\xi$, $\kappa_2$ and $\kappa_3$ take values distinct from those in Eq.~\eqref{eq:minimal_deltaCP}. In light of the weak evidence of $\delta_{CP}\sim-\pi/2$ from T2K experiment~\cite{Abe:2015awa}, we find that maximal Dirac CP phase $\delta_{CP}=\pm\pi/2$ would necessitate the following relation
\begin{equation}
\label{eq:maximal_deltaCP}\cos^2 \frac{\kappa_3-\kappa_2}{2}\tan2\xi=-\tan2\rho,\qquad
\kappa_3-\kappa_2\neq n_1\pi,\quad
\rho\neq\frac{n_2\pi}{2}, \quad n_{1,2}\in\mathbb{N}\,.
\end{equation}
If the permutation matrix $P$ takes a value different from Eq.~\eqref{eq:P_example}, the analytical expression of $\tan\delta_{CP}$ would differ from that in Eq.~\eqref{eq:mixing_para} so that the corresponding condition for maximal $\delta_{CP}$ would be distinct from Eq.~\eqref{eq:maximal_deltaCP}. This implies that the condition for $\cos\delta_{CP}=0$ depends on the column arrangement $P$.

In exactly a similar way, the necessary and sufficient conditions for trivial Majorana CP phase $\alpha_{21}$ are determined to be
\begin{equation}
\kappa_3=\kappa_2+n\pi,~~\mathrm{or}~~ \rho=n\pi/2,~~\mathrm{or}~~\rho=-\xi+n\pi~~\mathrm{with}~~n\in\mathbb{N}\,.
\end{equation}
The corresponding predictions for the mixing parameters are listed in Table~~\ref{tab:minimal_deltaCP_alpha_21}. The condition of maximal $\alpha_{21}$ is a bit complex as follows
\begin{equation}
a\cos^4\frac{\kappa_2-\kappa_3}{2}+b\cos^2\frac{\kappa_2-\kappa_3}{2}+c=0\,,
\end{equation}
where
\begin{eqnarray}
a&=&2\sin^22\xi\sin^22\rho,\nn\\
b&=&1+\sin4\xi\sin4\rho+\cos4\xi\left(1-2\cos4\rho\right),\nn\\
c&=&-2\cos^22\xi\sin^22\rho\,.
\end{eqnarray}
Now we turn to another Majorana phase. The redefined Majorana phase $\alpha^{\prime}_{31}$ would be conserved if
\begin{equation}
\tan\rho\tan\theta=\frac{\sin\frac{\kappa^{\prime}_2}{2}}{\sin\frac{\kappa^{\prime}_3}{2}}\quad\mathrm{or}\quad \frac{\cos\frac{\kappa^{\prime}_2}{2}}{\cos\frac{\kappa^{\prime}_3}{2}}\,.
\end{equation}
Maximal $\alpha^{\prime}_{31}$ requires the following equations be fulfilled,
\begin{equation}
\tan\rho\tan\theta=\frac{\sin\left(\frac{\pi}{4}+\frac{\kappa^{\prime}_2}{2}\right)}{\sin\left(\frac{\pi}{4}+\frac{\kappa^{\prime}_3}{2}\right)},\quad\mathrm{or}\quad \frac{\cos\left(\frac{\pi}{4}+\frac{\kappa^{\prime}_2}{2}\right)}{\cos\left(\frac{\pi}{4}+\frac{\kappa^{\prime}_3}{2}\right)}\,.
\end{equation}
Notice that the above conditions for minimal or maximal $\alpha_{21}$ and $\alpha^{\prime}_{31}$ vary with the permutation matrix $P$.

\section{\label{sec:K4_finite}Induced flavor symmetry arising from finite groups}

Under the assumption of Majorana neutrinos, the residual flavor symmetry of the neutrino mass matrix should be a Klein group, i.e. $G_{\nu}=\left\{1, G_1, G_2, G_3\right\}$ with $G^2_{i}=1$ and $G_{i}G_{j}=G_{k}$ for $i\neq j\neq k$. If both remnant flavor symmetries $G_{\nu}$ in the neutrino sector and $G_{l}$ in the charged lepton sector originate from a finite flavor symmetry group $G_{f}$, then $G_f$ would be strongly constrained. In the basis where the neutrino mass matrix $m_{\nu}$ is diagonal, the residual flavor symmetry transformation $G_{1,2,3}$ are given by
\begin{eqnarray}
G_1=\text{diag}(1,-1,-1)\,, \qquad G_2=\text{diag}(-1,1,-1)\,, \qquad G_3=\text{diag}(-1,-1,1)\,.
\label{Gnu}
\end{eqnarray}
The three-dimensional representation matrix of any element $g$ of $G_f$ can only be of the following form~\cite{Fonseca:2014koa}:
\begin{eqnarray}
|g|=\left(\begin{array}{ccc}
0                    ~&~   \frac{1}{\sqrt{2}}       ~&~   \frac{1}{\sqrt{2}}    \\
\frac{1}{\sqrt{2}}   ~&~     \frac{1}{2}           ~&~   \frac{1}{2}           \\
\frac{1}{\sqrt{2}}   ~&~      \frac{1}{2}           ~&~   \frac{1}{2}
\end{array}
\right)\,,\quad
\left(\begin{array}{ccc}
\frac{1}{2}      ~&~   \frac{\sqrt{5}-1}{4}    ~&~   \frac{\sqrt{5}+1}{4}    \\
\frac{\sqrt{5}+1}{4}    ~&~     \frac{1}{2}           ~&~   \frac{\sqrt{5}-1}{4}      \\
\frac{\sqrt{5}-1}{4}    ~&~  \frac{\sqrt{5}+1}{4}     ~&~   \frac{1}{2}
\end{array}
\right)\,,\quad\mathrm{or}\quad
\left(\begin{array}{ccc}
1   ~&~     0         ~&~       0       \\
0   ~&~   \cos\theta  ~&~   \sin\theta  \\
0   ~&~   \sin\theta  ~&~   \cos\theta
\end{array}
\right)
\label{eq:T235}
\end{eqnarray}
up to independent permutations of rows and columns, where $\theta/\pi$ is a rational number, and the abstract element $g$ and its representation matrix have been denoted by the same notation. Note that $G_{1,2,3}$ belong to the last category with $\theta=0$. In order to obtain a viable lepton mixing matrix in the experimentally preferred $3\sigma$ ranges, the residual flavor symmetry $G_{l}$ in the charged lepton sector should be generated by a single matrix $T$ with~\cite{Fonseca:2014koa}
\begin{equation}
T=\left(\begin{array}{ccc}
0                  ~&~   \frac{1}{\sqrt{2}}       ~&~   \frac{1}{\sqrt{2}}    \\
\frac{1}{\sqrt{2}}   ~&~     -\frac{1}{2}           ~&~   \frac{1}{2}           \\
\frac{1}{\sqrt{2}}   ~&~      \frac{1}{2}           ~&~   -\frac{1}{2}
\end{array}
\right)
\left(\begin{array}{ccc}
e^{i\zeta_1}      ~&~        0              ~&~       0            \\
0                 ~&~     e^{i\zeta_2}      ~&~       0            \\
0                 ~&~        0              ~&~  -e^{i\zeta_2}
\end{array}
\right)\,,
\end{equation}
where both $\zeta_1/\pi$ and $\zeta_2/\pi$ are rational numbers.
The charged lepton diagonalization matrix $U_{l}$ is determined by the diagonalization of $T$, and consequently we have
\begin{equation}
\label{eq:Ul}U_l=\frac{1}{\sqrt{3}}\left(
\begin{array}{ccc}
1     ~&~   1    ~&~ 1\\
-\sqrt{2}e^{-i\frac{\zeta_2^\prime}{2}}\cos\frac{\zeta_2^\prime}{6}     ~&~\sqrt{2}e^{-i\frac{\zeta_2^\prime}{2}}\cos\left(\frac{\zeta_2^\prime}{6}+\frac{\pi}{3} \right)   ~&~ \sqrt{2}e^{-i\frac{\zeta_2^\prime}{2}}\cos\left(\frac{\zeta_2^\prime}{6}-\frac{\pi}{3}\right)\\
\sqrt{2}i e^{-i\frac{\zeta_2^\prime}{2}}\sin\frac{\zeta_2^\prime}{6}  ~&~
-\sqrt{2}i e^{-i\frac{\zeta_2^\prime}{2}}\sin\left(\frac{\zeta_2^\prime}{6}+\frac{\pi}{3}\right)   ~&~-\sqrt{2}i e^{-i\frac{\zeta_2^\prime}{2}}\sin\left(\frac{\zeta_2^\prime}{6}-\frac{\pi}{3}\right)
\end{array}
\right)\,,
\end{equation}
with $U^{\dagger}_{l}TU_{l}=\mathrm{diag}(-e^{i(\zeta_1+2\zeta_2)/3}, e^{i(\pi+\zeta_1+2\zeta_2)/3}, e^{i(-\pi+\zeta_1+2\zeta_2)/3})$, where $\zeta^{\prime}_2\equiv\zeta_2-\zeta_1$ is defined for simplicity. In order to predict the values of the Majorana CP phases, we impose CP symmetry onto the theory. In the present $m_{\nu}$ diagonal basis, the residual CP transformations of the neutrino mass matrix can only be diagonal, i.e.
\begin{eqnarray}
X_1&=&\text{diag}(e^{i\alpha_1}, -e^{i\alpha_2}, -e^{i\alpha_3}), \qquad X_2=\text{diag}(-e^{i\alpha_1}, e^{i\alpha_2}, -e^{i\alpha_3}),\nn\\
X_3&=&\text{diag}(-e^{i\alpha_1}, -e^{i\alpha_2}, e^{i\alpha_3}), \qquad X_4=\text{diag}(e^{i\alpha_1}, e^{i\alpha_2}, e^{i\alpha_3})\,,\label{eq:residual_CP_finite}
\end{eqnarray}
where $\alpha_1$, $\alpha_2$ and $\alpha_3$ are real parameters. The same set of residual CP transformations can be obtained by solving the equation $X_iG^{\ast}_{j}X^{-1}_{i}=G_{j}$ given in Eq.~\eqref{eq:XGcond}. Moreover, obviously the residual flavor symmetry $G_{1,2,3}$ can be generated by successively performing two CP transformations $X_{i}X^{\ast}_j$ as shown in Eq.~\eqref{eq:CP_flavor}. Now we shall proceed to study the constraints on the phases $\alpha_{1}$, $\alpha_{2}$ and $\alpha_{3}$ provided that $G_{f}$ is a finite group. If we firstly perform a CP transformation $X_4$, subsequently a flavor symmetry transformation $T$, and eventually an inverse CP transformation $X^{-1}_4$, the theory is still invariant.
The total effect of this series of transformations should
amount to a flavor symmetry transformation,
\begin{equation}
T^{\prime}\equiv  X_4 T^{*} X_4^{-1}\in G_{f}\,,
\end{equation}
which is the so-called consistency condition~\cite{Feruglio:2012cw,Holthausen:2012dk,Chen:2014tpa}. Therefore $T^{\prime}T$ is an element of $G_{f}$ as well, and the absolute value of $T^{\prime}T$ is
\begin{equation}
|T^{\prime}T|=\left(
\begin{array}{ccc}
A&B&B \\
B&D&C \\
B&C&D
\end{array}
\right)\,,
\end{equation}
where
\begin{eqnarray}
A&=&|\sin\frac{\alpha_3-\alpha_2}{2}|\,,\nn\\
B&=&\frac{1}{\sqrt{2}}|\cos\frac{\alpha_3-\alpha_2}{2}|\,,\nn\\
C&=&\frac{1}{2\sqrt{2}}\sqrt{3-\cos(\alpha_3-\alpha_2)-2\cos(\alpha_2-\alpha_1+\zeta^{\prime}_2)+2\cos(\alpha_3-\alpha_1+\zeta^{\prime}_2)}\,,\nn\\
D&=&\frac{1}{2\sqrt{2}}\sqrt{3-\cos(\alpha_3-\alpha_2)-2\cos(\alpha_3-\alpha_1+\zeta^{\prime}_2)+2\cos(\alpha_2-\alpha_1+\zeta^{\prime}_2)}\,.
\end{eqnarray}
If the flavor symmetry group generated by $G_{1,2,3}$, $T$ and $T^{\prime}$ is finite, $|T^{\prime}T|$ has to be of the forms in Eq.~\eqref{eq:T235} up to row and column permutations. As a result, $B$ can only be equal to $0$, $\frac{1}{2}$ or $\frac{1}{\sqrt{2}}$. Then we can derive that
\begin{equation}
\alpha_3=\alpha_2,\quad \mathrm{or} \quad \alpha_3=\alpha_2\pm\pi\,.
\end{equation}
Since both solutions $\alpha_3=\alpha_2$ and $\alpha_3=\alpha_2\pm\pi$ give rise to the same set of residual CP transformations $X_{1,2,3,4}$ except an inessential overall ``$-1$'' factor, we shall choose $\alpha_3=\alpha_2$ in the following without loss of generality. The eigenvalues of $T^{\prime\dagger}T$ are $e^{i(\alpha_1-\alpha_2+2\zeta_1)}$, $e^{2i\zeta_2}$ and $e^{i(\alpha_2-\alpha_1+2\zeta_2)}$. Once the flavor symmetry group $G_{f}$ is finite,  the order of $T^{\prime\dagger}T$ must be finite as well. As a consequence, $\alpha^{\prime}_{2}\equiv\alpha_{2}-\alpha_1$ a rational multiple of $\pi$. The neutrino mass matrix $m_{\nu}$ is constrained by the CP symmetry as
\begin{equation}
X_i^Tm_{\nu}X_i=m^{*}_\nu\,.
\end{equation}
Therefore $m_{\nu}$ is fixed to be
\begin{equation}
m_{\nu}=\text{diag}(m_1e^{-i\alpha_1}, m_2e^{-i\alpha_2}, m_3e^{-i\alpha_2})\,,
\end{equation}
where $m_1$, $m_2$ and $m_3$ are real. The unitary diagonalization matrix $U_{\nu}$, which fulfills $U^T_{\nu}m_{\nu}U_{\nu}=\text{diag}(|m_1|,|m_2|,|m_3|)$, is given by
\begin{equation}
\label{eq:Unu}U_{\nu}=\left(\begin{array}{ccc}
e^{i\frac{\alpha_1}{2}}&0&0 \\
0&e^{i\frac{\alpha_2}{2}}&0    \\
0&0&e^{i\frac{\alpha_2}{2}}
\end{array}
\right)K\,,
\end{equation}
where $K$ is a diagonal matrix with entries $\pm1$ or $\pm i$ which encode the CP parity of the neutrino states and it makes the light neutrino masses positive. Notice that $U_{\nu}$ is determined up to permutations and phases of its column vectors, since the order of the neutrino masses is undefined in this approach. Compared with the scenario with only flavor symmetry, the unitary matrix $U_{\nu}$ can be any diagonal phase matrix.

Combining the unitary transformations $U_{l}$ in Eq.~\eqref{eq:Ul} and $U_{\nu}$ in Eq.~\eqref{eq:Unu}, we can obtain the prediction for the lepton mixing matrix:
\begin{equation}
\label{eq:UPMNS_finite}U_{PMNS}=\frac{1}{\sqrt{3}}\left(\begin{array}{ccc}
-\sqrt{2}\cos\frac{\zeta_2^\prime}{6}  &  ~1~     & \sqrt{2}\sin\frac{\zeta_2^\prime}{6} \\
\sqrt{2}\cos\left(\frac{\zeta_2^\prime}{6}+\frac{\pi}{3}\right)   &
~1~&-\sqrt{2}\sin\left(\frac{\zeta_2^\prime}{6}+\frac{\pi}{3}\right)\\
\sqrt{2}\cos\left(\frac{\zeta_2^\prime}{6}-\frac{\pi}{3}\right)  &
~1~ & -\sqrt{2}\sin\left(\frac{\zeta_2^\prime}{6}-\frac{\pi}{3}\right)
\end{array}\right)
\left(\begin{array}{ccc}
e^{i\frac{\zeta_2^\prime+\alpha_2^\prime}{2}} & 0 & 0   \\
0  &  1  &  0    \\
0  &  0  &  ~-ie^{i\frac{\zeta_2^\prime+\alpha_2^\prime}{2}}
\end{array}\right)K\,.
\end{equation}
Permuting the rows and columns of this mixing matrix while keeping the vector $(1, 1, 1)^{T}/\sqrt{3}$ in the second column, we find the resulting PMNS matrix can be obtained from Eq.~\eqref{eq:UPMNS_finite} by redefinition of the parameters $\zeta^{\prime}_2$ and $\alpha^{\prime}_2$. On the other hand, we can also turn to the charged lepton diagonal basis (i.e. the $T$ diagonal basis) by performing the similarity transformation $U_{l}$ of Eq.~\eqref{eq:Ul}. The residual CP transformations $X_i$ in Eq.~\eqref{eq:residual_CP_finite} become $U^{\dagger}_{l}X_iU^{\ast}_{l}$ which are characterized by the parameter values,
\begin{eqnarray}
\varphi&=&\arccos\sqrt{\frac{1}{3}},\quad \phi=\frac{\pi}{4},\quad \rho=-\frac{\zeta_2^\prime}{6}, \quad \xi=0,\nn\\
\kappa_1&=&\alpha_1,\quad \kappa_2=\pi+\alpha_2+\zeta_2^\prime,\quad
\kappa_3=\alpha_2+\zeta_2^\prime\,.
\end{eqnarray}
Then the PMNS matrix can be easily obtained via the general formula of Eq.~\eqref{eq:UPMNS_final}. We can further straightforwardly read out the lepton mixing parameters as follows
\begin{eqnarray}
\sin^2\theta_{13}&=&\frac{2}{3}\sin^2\frac{\zeta_2^\prime}{6},\quad \sin^2\theta_{12}=\frac{1}{2+\cos\frac{\zeta_2^\prime}{3}},\quad
\sin^2\theta_{23}=\frac{1}{2}\left(1+\frac{\sqrt{3}\sin\frac{\zeta_2^\prime}{3}}
{2+\cos\frac{\zeta_2^\prime}{3}}\right),\nn\\
\tan\delta_{CP}&=&0,\quad \tan\alpha_{21}=-\tan\left(\zeta_2^\prime+\alpha_2^\prime\right),\quad
\tan\alpha_{31}=0\,.\label{eq:mixing_parames_K4}
\end{eqnarray}
It is remarkable that both Dirac phase $\delta_{CP}$ and the Majorana phase $\alpha_{31}$ are trivial, and another Majorana phase $\alpha_{21}=-\zeta_2^\prime-\alpha_2^\prime$ or $\alpha_{21}=\pi-\zeta_2^\prime-\alpha_2^\prime$ is a rational angle. Note that these predictions for the CP phases are independent of the structure of $G_{f}$. These results are exactly found to be true in the context of $G_{f}=\Delta(6n^2)$ flavor symmetry with generalized CP, as will be discussed in the following section. We can see that the three mixing angles depend on only one parameter $\zeta_2^\prime$ with period of $6\pi$. Without loss of generality, the fundamental interval of the parameter $\zeta_2^\prime$ can be taken to be $[-3\pi, 3\pi)$. Imposing the present experimentally favored $3\sigma$ experiment value~\cite{Gonzalez-Garcia:2014bfa}, $\zeta_2^\prime/\pi$ is a rational number satisfying $0.322\le\pm\zeta_2^\prime/\pi\le0.373$. Moreover, the three mixing angles are strongly correlated with each other,
\begin{equation}
3\cos^2\theta_{13}\sin^2\theta_{12}=1,\qquad \sin^2\theta_{23}=\frac{1}{2}\pm\frac{1}{2}\tan\theta_{13}\sqrt{2-\tan^2\theta_{13}}\,.
\end{equation}
Given the $3\sigma$ range $0.0188\leq\sin^2\theta_{13}\leq0.0251$, we obtain $0.340\leq\sin^2\theta_{12}\leq0.342$, $0.387\leq\sin^2\theta_{23}\leq0.403$ or $0.597\leq\sin^2\theta_{23}\leq0.613$. Note that the atmospheric angle $\theta_{23}$ deviates from maximal mixing. These predictions for $\theta_{12}$ and $\theta_{23}$ can be tested by forthcoming neutrino oscillation experiments such as JUNO~\cite{JUNO_exp}, LBNE~\cite{Adams:2013qkq} and LBNO~\cite{::2013kaa}. Moreover, dedicated long baseline experiments LBNE~\cite{Adams:2013qkq}, LBNO~\cite{::2013kaa} and Hyper-Kamiokande~\cite{Abe:2011ts} aim to measure the Dirac CP phase $\delta_{CP}$, the present scenario will be ruled out if the signal of leptonic CP violation will be detected.

\begin{figure}[t!]
\begin{center}
\includegraphics[width=0.99\linewidth]{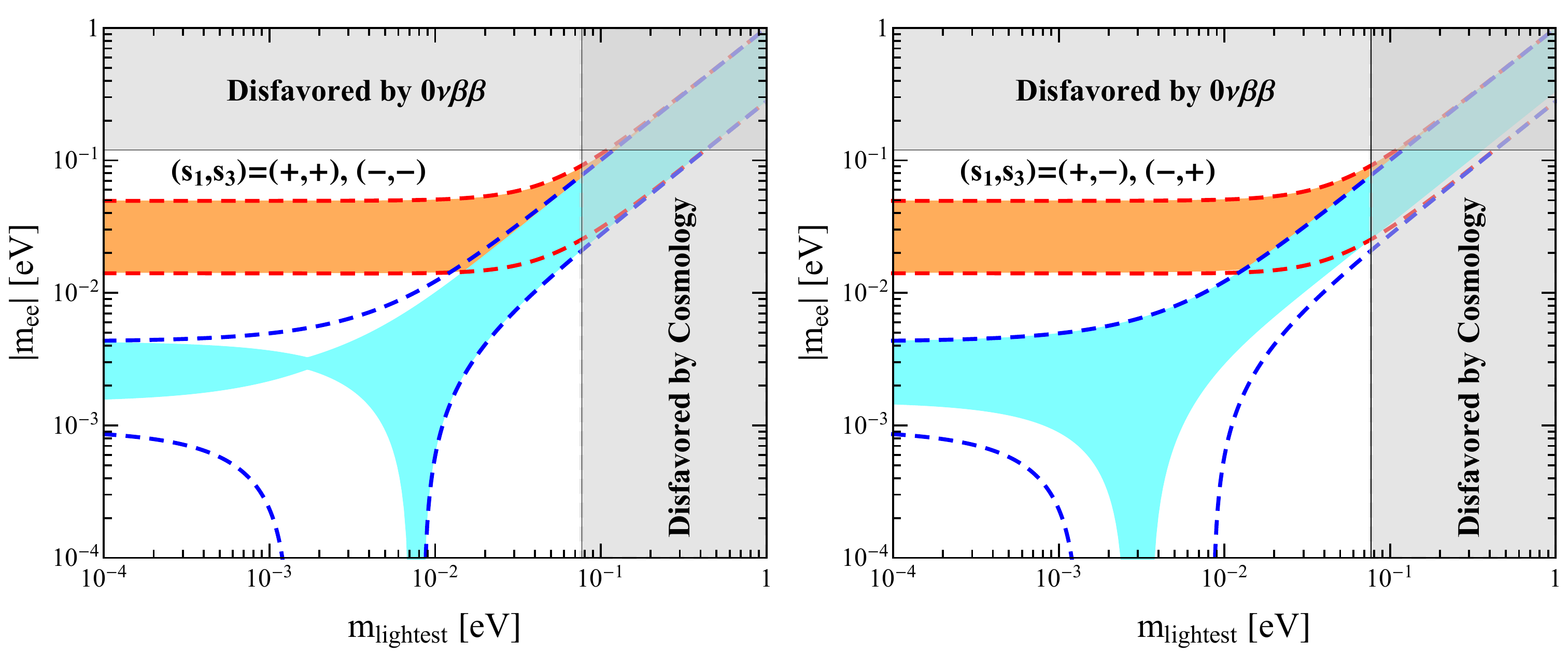}
\caption{\label{fig:mee}The prediction for the $0\nu\beta\beta$ decay effective mass if the Klein flavor symmetry generated by the residual CP symmetry originates from a finite flavor symmetry group. In this case, the PMNS matrix is given in Eq.~\eqref{eq:UPMNS_finite}, both $\delta_{CP}$ and $\alpha_{31}$ are predicted to be trivial, and $\alpha_{21}$ is a rational angle. The blue dashed lines and the red dashed lines indicate the currently allowed $3\sigma$ regions for normal ordering and inverted ordering mass spectrum respectively~\cite{Gonzalez-Garcia:2014bfa}. The cyan and orange areas are theoretical predictions when $\alpha^{\prime}_2$ freely varies in the interval $[0, 2\pi]$ and $\zeta^{\prime}_2$ in the viable range $0.322\pi\leq\pm\zeta^{\prime}_2\leq0.373\pi$. Measurements of EXO-200~\cite{Auger:2012ar,Albert:2014awa} in combination with KamLAND-ZEN~\cite{Gando:2012zm} give rise to the upper bound of $|m_{ee}|<0.120$ eV. The upper limit on the mass of the lightest neutrino is derived from the latest Planck result $m_1+m_2+m_3<0.230$ eV at $95\%$ level~\cite{Ade:2013zuv}.}
\end{center}
\end{figure}

The neutrinoless double beta ($0\nu\beta\beta$) decay is the only feasible experiment which has the potential of establishing the Majorana nature of massive neutrinos. The dependence of the decay rate on the mixing parameters is specified by the effective Majorana mass,
\begin{equation}
\label{eq:mee}\left|m_{ee}\right|=\left|m_1\cos^2\theta_{12}\cos^2\theta_{13}+m_2\sin^2\theta_{12}\cos^2\theta_{13}e^{i\alpha_{21}}+m_3\sin^2\theta_{13}e^{i(\alpha_{31}-2\delta_{CP})}\right|\,.
\end{equation}
For the predicted mixing pattern in Eq.~\eqref{eq:UPMNS_finite}, we have
\begin{equation}
\left|m_{ee}\right|=\frac{1}{3}\left|2s_1m_1\cos^2\frac{\zeta^{\prime}_2}{6}+m_2e^{-i(\zeta^{\prime}_2+\alpha^{\prime}_2)}-2s_3m_3\sin^2\frac{\zeta^{\prime}_2}{6}\right|\,,
\end{equation}
where $s_{1,3}=\pm1$ arises from the ambiguity of the phase matrix $K$. The parameter $\alpha^{\prime}_2$ freely varies in the region of $0\leq\alpha^{\prime}_2\leq2\pi$ and $\zeta^{\prime}_2$ is scattered in the viable ranges of $0.322\pi\leq\pm\zeta^{\prime}_2\leq0.373\pi$. The resulting predictions for the effective mass $|m_{ee}|$ are plotted in Fig.~\ref{fig:mee}, where the $3\sigma$ uncertainties of the mass-squared splittings $\Delta m^2_{21}$ and $\Delta m^2_{3\ell}$ with $\ell=1$ for normal ordering and $\ell=2$ for inverted ordering are included~\cite{Gonzalez-Garcia:2014bfa}. We see that all possible values of $|m_{ee}|$ allowed by experimental data at $3\sigma$ level can be achieved in case of inverted ordering mass spectrum.

\section{\label{sec:Delta6n2}Example with $G_{f}=\Delta(6n^2)$}

In what follows, we shall substantiate the statement that the lepton mixing matrix must be the trimaximal pattern in Eq.~\eqref{eq:UPMNS_finite} if the residual flavor symmetry generated by the residual CP transformations originates from a finite group $G_{f}$. As a example and a further check to our general results, we consider the case of $G_{f}=\Delta(6n^2)$. $\Delta(6n^2)$ is a series of non-abelian finite subgroup of $SU(3)$, and it can be generated by four generators $a$, $b$, $c$ and $d$ which fulfill~\cite{Escobar:2008vc}
\begin{eqnarray}
&a^3=b^2=(ab)^2=1,\nn\\
&c^{n}=d^{n}=1,\quad cd=dc,\nn\\
&aca^{-1}=c^{-1}d^{-1},\quad ada^{-1}=c,\quad bcb^{-1}=d^{-1},\quad bdb^{-1}=c^{-1}\,.
\end{eqnarray}
$\Delta(6n^2)$ group has $2n-2$ three dimensional irreducible representations denoted by $\mathbf{3}_{l,k}$, and representation matrices of the generators are given by
\begin{eqnarray}
a=\left(\begin{array}{ccc}
0 &  1  &  0 \\
0 &  0  &  1 \\
1 &  0  &  0
\end{array}
\right),\quad b=(-1)^{l}\left(\begin{array}{ccc}
0  & 0 & 1  \\
0  & 1 & 0 \\
1  & 0 & 0
\end{array}
\right),\quad c=\left(\begin{array}{ccc}
\eta^{k}  &  0  &  0 \\
0   &  \eta^{-k}  &  0 \\
0   &  0  &  1
\end{array}\right),\quad d=\left(\begin{array}{ccc}
1   &  0   &  0 \\
0   &  \eta^{k}  &  0  \\
0   &   0    &  \eta^{-k}
\end{array}\right)\,,
\end{eqnarray}
where $\eta=e^{2\pi i/n}$, $l=1, 2$ and $k=1, 2, \ldots, n-1$. We see that the set of all matrices describing distinct triplet representations $\mathbf{3}_{l,k}$ is the same up to a possible overall ``$-1$'' factor. As a consequence, it is sufficient to focus on a single three-dimensional representation in the discussion of the mixing patterns. We shall assign the three generations of left-handed leptons to the three-dimensional representation $\mathbf{3}_{1,1}$ in the following. To simplify the notation, the abstract elements of $\Delta(6n^2)$ and their representation matrices have been denoted by the same symbol.

\begin{figure}[t!]
\begin{center}
\includegraphics[width=0.98\linewidth]{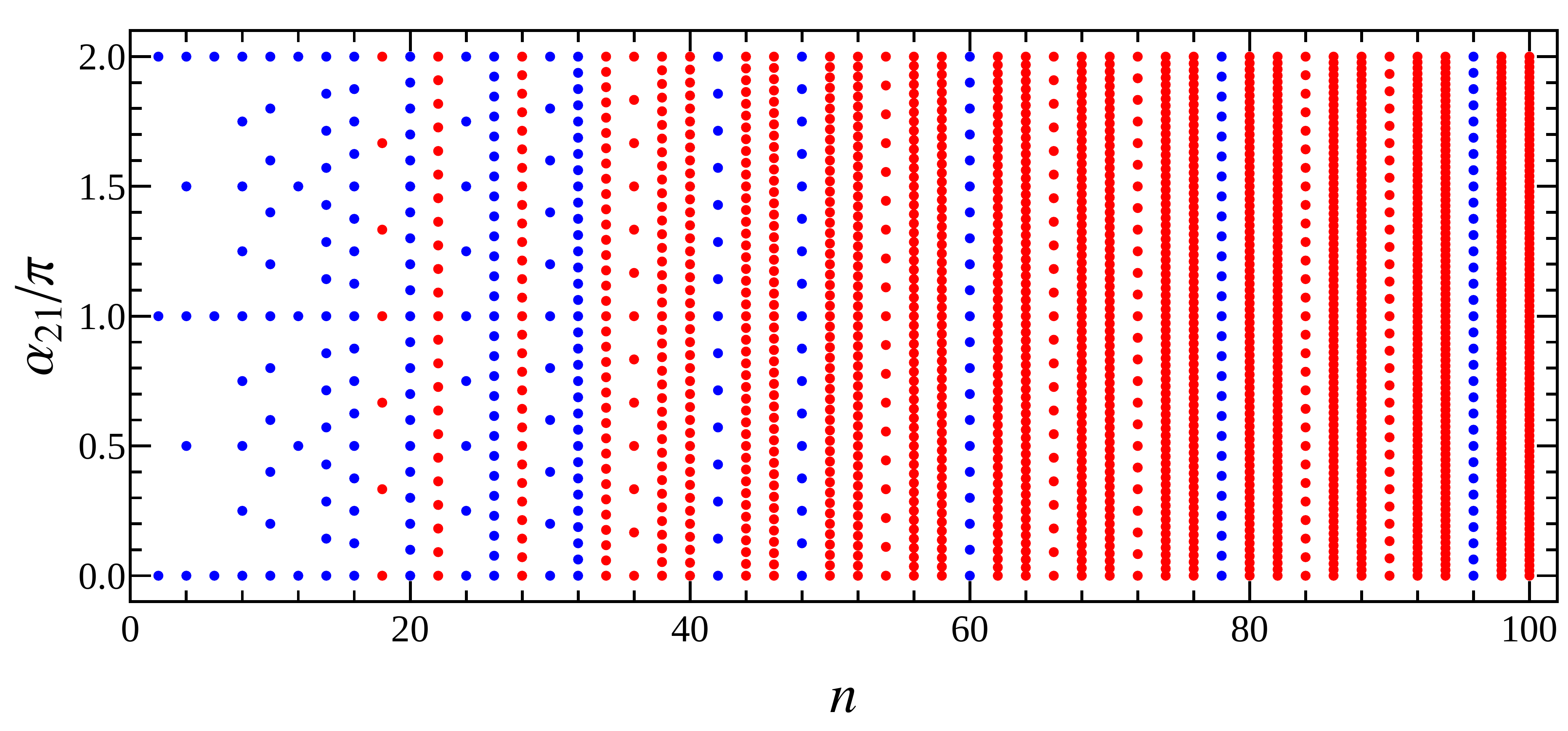}
\caption{\label{fig:alpha21}The possible values of the Majorana phase $\alpha_{21}$ predicted by CP symmetry if the residual Klein flavor symmetry generated by the residual CP originates from the $\Delta(6n^2)$ group with even $n$. The red points denote the predictions for $\alpha_{21}$, where the lepton mixing angles are required to lie in the experimentally favored $3\sigma$ intervals~\cite{Gonzalez-Garcia:2014bfa}. The blue points represent that the measured values of $\theta_{13}$ can not be accommodated.}
\end{center}
\end{figure}

$\Delta(6n^2)$ as a flavor symmetry group has been studied comprehensively~\cite{King:2013vna,King:2014rwa,Hagedorn:2014wha,Ding:2014ora}. It is found that CP symmetry can be consistently defined in the context of $\Delta(6n^2)$ family symmetry if $n$ is not divisible by 3 or the fields transforming as two dimensional irreducible representation $\mathbf{2_2}$, $\mathbf{2_3}$ or $\mathbf{2_4}$ are not present in a concrete model~\cite{Ding:2014ora}. In particular, the generalized CP transformations are of the same form as the flavor symmetry transformation in the working basis~\cite{Ding:2014ora}.
Phenomenologically viable lepton flavor mixing can be obtained if $\Delta(6n^2)$ is broken down to $G_{l}=Z^{a}_3\equiv\left\{1, a, a^2\right\}$ in the charged lepton sector and to Klein subgroup $G_{\nu}=K^{(c^{n/2}, abc^{\gamma})}_4\equiv\left\{1, c^{n/2}, abc^{\gamma}, abc^{\gamma+n/2}\right\}$ in the neutrino sector~\cite{King:2013vna,King:2014rwa,Yao:2015dwa}, where $\gamma=0, 1, \ldots, n-1$ and $n$ should be even. In this work, we propose to start from CP symmetry rather than flavor symmetry. The remnant flavor symmetry $K^{(c^{n/2}, abc^{\gamma})}_4$ can be generated if impose the following CP transformations
\begin{equation}
\label{eq:GCP_delta6n2}X_1=c^{s}d^{2(s+\gamma)},\quad X_2=abc^{s+\gamma}d^{2(s+\gamma)},\quad X_3=abc^{s+\gamma+n/2}d^{2(s+\gamma)},\quad X_4=c^{s+n/2}d^{2(s+\gamma)}\,,
\end{equation}
where $s=0, 1, \ldots, n-1$. Note that the remnant symmetry $G_{l}=Z^{a}_3$ can be generated by the CP transformations $X_{l}=b,\;ab,\;a^2b$. The generator $a$ and the hermitian combination $m^{\dagger}_lm_l$ are diagonalized by the same unitary transformation $U_{l}$ with
\begin{equation}
U_{l}=\frac{1}{\sqrt{3}}\left(\begin{array}{ccc}
1   &   \omega^2   &  \omega  \\
1   &   \omega     &   \omega^2  \\
1   &    1       &  1
\end{array}
\right),\quad U^{\dagger}_{l}aU_{l}=\mathrm{diag}\left(1, \omega^2, \omega\right)\,,
\end{equation}
where $\omega=e^{2\pi i/3}$. We now apply our method to this specific example. Firstly we perform a change of basis with the unitary matrix $U_{l}$, such that the charged lepton mass matrix would be diagonal. The postulated residual CP transformations in Eq.~\eqref{eq:GCP_delta6n2} transform into $U^{\dagger}_{l}X_iU^{\ast}_l$, they can be parameterized in the manner shown in section~\ref{sec:constructing_PMNS_matrix}, and the parameter values are given by
\begin{eqnarray}
&&\qquad~\varphi=\arccos\frac{1}{\sqrt{3}},\quad \phi=\frac{\pi}{4},\quad  \rho=-\frac{\gamma\pi}{n},\quad \xi=0,\nn\\ &&\kappa_1=-\frac{4(s+\gamma)\pi}{n},\quad \kappa_2=\frac{2(s+\gamma)\pi}{n},\quad \kappa_3=\pi+\frac{2(s+\gamma)\pi}{n}\,.
\end{eqnarray}
Using our formula for the PMNS matrix in Eq.~\eqref{eq:UPMNS_final}, we can construct the mixing matrix as
\begin{equation}
U_{PMNS}=\frac{1}{\sqrt{3}}\left(\begin{array}{ccc}
-\sqrt{2}\cos\frac{\gamma\pi}{n}   &   ~1~   &   \sqrt{2}\sin\frac{\gamma\pi}{n}  \\
\sqrt{2}\cos\left(\frac{\gamma\pi}{n}+\frac{\pi}{3}\right)  &  ~1~   &  -\sqrt{2}\sin\left(\frac{\gamma\pi}{n}+\frac{\pi}{3}\right)  \\
\sqrt{2}\cos\left(\frac{\gamma\pi}{n}-\frac{\pi}{3}\right)   &   ~1~   &  -\sqrt{2}\sin\left(\frac{\gamma\pi}{n}-\frac{\pi}{3}\right)
\end{array}
\right)\left(\begin{array}{ccc}
e^{\frac{3i\pi(s+\gamma)}{n}}  &  0   &  0\\
0  &  1   &  0 \\
0  &  0   & -ie^{\frac{3i\pi(s+\gamma)}{n}}
\end{array}
\right)K\,,
\end{equation}
which is of the same form as our general result of Eq.~\eqref{eq:UPMNS_finite}. As a consequence, the lepton mixing parameters are given by Eq.~\eqref{eq:mixing_parames_K4} with $\zeta^{\prime}_2=\frac{6\gamma\pi}{n}$ and $\alpha^{\prime}_2=\frac{6s\pi}{n}$. In particularly, both $\delta_{CP}$ and $\alpha_{31}$ are conserved, and $\alpha_{21}=-\frac{6(s+\gamma)\pi}{n}$ or $\alpha_{21}=\pi-\frac{6(s+\gamma)\pi}{n}$. The same results are obtained in Ref.~\cite{King:2014rwa}. We show the possible values of the Majorana phase $\alpha_{21}$ for each $\Delta(6n^2)$ group of even $n$ in Fig.~\ref{fig:alpha21}. For large $n$, we see that the predictions for $\alpha_{21}$ densely fill the whole range of $[0, 2\pi]$. For small values of $n$, the measured value of reactor angle $\theta_{13}$ can not be generated, and the corresponding values of $\alpha_{21}$ are plotted in blue color. The red points denote the possible values of $\alpha_{21}$ when the experimentally favored $3\sigma$ ranges of the mixing angles are taken into account.

\section{\label{sec:conclusions}Conclusions}

CP symmetry is a more general framework than the flavor symmetry in constraining the lepton flavor mixing, since flavor symmetry can be generated by performing two CP transformations. Compared with flavor symmetry, all mixing parameters in particular the Majorana phases can be predicted by CP symmetry. In the charged lepton diagonal basis, a generic Majorana neutrino mass matrix has four remnant CP transformations which are determined by the experimentally measured PMNS matrix. Note that only three of the four remnant CP transformations are independent. Conversely the lepton flavor mixing PMNS matrix can be constructed from the postulated remnant CP symmetries. If only one CP transformation is preserved by the neutrino mass matrix, the PMNS matrix would be determined up to an arbitrary real orthogonal matrix. If two CP transformations are preserved in the neutrino sector, the PMNS matrix would  depend on a single real free parameters besides the parameters characterizing the remnant CP transformations. The explicit form of the PMNS matrix for both one and two remnant CP has been derived in our previous work~\cite{Chen:2014wxa}.

In the present work, we have considered the scenario that four CP transformations out of the original CP symmetry at high energy scale are conserved by the neutrino mass matrix. A remnant Klein four flavor symmetry would be generated in this case. We firstly present the most general parameterization of the four remnant CP transformations, and then the reconstruction formula for the PMNS matrix is derived. We see that the PMNS matrix including the Majorana phases are completely fixed by the postulated four remnant CP transformations. From the explicit form of the PMNS matrix, the necessary and sufficient condition for conserved Dirac CP violating phases is determined to be $\xi=0,~\pi/2,~\pi,~3\pi/2$, $\kappa_3=\kappa_2$ or $\kappa_3=\kappa_2\pm\pi$. In the same fashion, we find the conditions for maximal $\delta_{CP}$, minimal Majorana phases and maximal Majorana phases for the column permutation $P$ given by Eq.~\eqref{eq:P_example}.

Furthermore, we discuss the situation that the Klein four flavor symmetry induced by the remnant CP originates from a finite flavor symmetry group. It turns out that the lepton flavor mixing would be strongly constrained. The phenomenologically viable PMNS matrix can only take the trimaximal form, and it depends on two rational angles $\zeta^{\prime}_2$ and $\alpha^{\prime}_2$, as shown in Eq.~\eqref{eq:UPMNS_finite}. As a consequence, the three lepton mixing angles are correlated with each other. Given the measured values of the reactor angle $\theta_{13}$, we have $\sin^2\theta_{12}\simeq0.341$, $\sin^2\theta_{23}\simeq0.395$ or $\sin^2\theta_{23}\simeq0.605$. Regarding the CP violating phases, both $\delta_{CP}$ and $\alpha_{31}$ are determined to be 0 or $\pi$ while another Majorana phase $\alpha_{21}$ can be any rational angle. These predictions can be tested by more precise neutrino oscillation experiments in near future. In addition, the corresponding predictions for the neutrinoless double beta decay are studied. The $3\sigma$ region of the effective mass $|m_{ee}|$ can be nearly reproduced in the case of inverted ordering neutrino mass spectrum. As a concrete example, we further consider the case that the induced Klein four flavor symmetry arises from the $\Delta(6n^2)$ flavor symmetry group. The PMNS matrix is really found to be the trimaximal pattern with $\zeta^{\prime}_2=6\gamma\pi/n$ and $\alpha^{\prime}_2=6s\pi/n$, where $\gamma, s=0, 1, 2,\ldots, n-1$. The above general results are confirmed.

\section*{Acknowledgements}

This work is supported by the National Natural Science Foundation of China under Grant Nos. 11275188 and 11179007.

\end{document}